\documentclass[nofootinbib,prl,twocolumn,amsmath,amssymb,aps]{revtex4-1}

\usepackage{graphicx}
\usepackage{dcolumn}
\usepackage{bm}
\usepackage{hyperref}
\usepackage{cleveref}
\usepackage{color}

\def\beq{\begin{equation}}
\def\eeq{\end{equation}}
\def\bea{\begin{align}}
\def\eea{\end{align}}

\def\bk{{\bf k}}
\def\bp{{\bf p}}
\def\bq{{\bf q}}

\def\la{\langle}
\def\ra{\rangle}

\def\calZ{\mathcal{Z}}
\def\calD{\mathcal{D}}

\def\calV{\mathcal{V}}

\def\e{\epsilon}

\def\nn{\nonumber}

\begin{document}

\title{Induced $p$-wave pairing in Bose-Fermi mixtures}

\author{Jami J. Kinnunen$^{1}$, Zhigang Wu$^{2}$ and Georg M.~Bruun$^{3}$}

\affiliation{$^{1}$Department of Applied Physics, Aalto University, FI-00076 Aalto, Finland}
\affiliation{$^{2}$Shenzhen Institute for Quantum Science and Engineering and Department of Physics, Southern University of Science and Technology, Shenzhen 518055, China}
\affiliation{$^{3}$Department of Physics and Astronomy, Aarhus University, DK-8000 Aarhus C, Denmark}

\date{\today}

\begin{abstract}
 Cooper pairing  caused by an  induced  interaction represents a paradigm in our description of fermionic superfluidity.  Here, 
 we present a strong coupling theory for the critical temperature of $p$-wave pairing between spin polarised fermions immersed in a Bose-Einstein condensate. 
 The fermions interact via the exchange of phonons in the condensate, and our self-consistent theory takes into account the full frequency/momentum dependence of 
 the resulting induced interaction. We demonstrate that both retardation and self-energy effects are important for obtaining a reliable value of the  
 critical temperature. Focusing on experimentally relevant systems, we perform a systematic analysis varying 
 the boson-boson and boson-fermion interaction strength as well as their masses, and identify the most suitable   system for realising a  $p$-wave superfluid. 
 Our results show that such a superfluid   indeed is experimentally within reach using light bosons mixed with  
 heavy fermions. 
\end{abstract}

\maketitle

The theory of Cooper pairing of electrons  due to an induced attractive interaction mediated via crystal phonons successfully explained the origin of superconductivity and
stands out as a highlight of quantum many-body physics~\cite{Schrieffer1983}. Interest in pairing 
with non $s$-wave symmetry began with the understanding of superfluidity in $^3$He~\cite{Vollhardt1990} and has 
increased further with the advent of systems such as the copper and 
iron based high temperature superconductors~\cite{Bednorz1986,Kamihara2008}, for which  many fundamental questions remain.  Ultracold atoms 
have emerged as a powerful platform to  explore such many-body physics, and the realization of strong $s$-wave pairing in 
a  Fermi gas was a landmark achievement~\cite{Regal2004,Zwierlein2004}. Pairing in these gases is however brought by a direct 
attractive interaction between the fermions, and so far no one has realised pairing via an induced interaction in cold atom systems. 

Spin-polarised fermions mixed with a Bose-Einstein condensate (BEC) represents a promising set-up for realising a $p$-wave superfluid caused by an induced interaction. 
Here,  the fermions gain an effective attraction through exchanging phonons in the BEC~\cite{Heiselberg2000,Bijlsma2000}.
 A very attractive feature of such a mediated $p$-wave interaction is that both its strength and  range can be tuned by changing the properties of the BEC. 
 Experimentally, much progress has been made recently on atomic mixtures and the list of trapped Bose-Fermi mixtures is already  
 long~\cite{Ferrier-Barbut2014,Roy2017,Yao2016,Zaccanti2006,Ospelkaus2006,Taglieber2008,Park2012,Heo2012,Vaidya2015,DeSalvo2017,Lous2018,Schafer2018}.
 Cooper pairing in atomic Bose-Fermi mixtures was originally predicted using weak coupling BCS theory~\cite{Efremov2002}, and 
 since then  several authors have considered the problem using theories with varying degree of sophistication~\cite{Matera2003,Bulgac2006,Suzuki2008,Bulgac2009,Matera2011}.
 It has furthermore  been predicted that topological $p$-wave superfluids can be realised in mixed dimensional Bose-Fermi mixtures~\cite{Wu2012,Okamoto2017}. However, a  strong coupling theory for the critical temperature of a three-dimensional
 $p$-wave superfluid  including the full energy and momentum dependent 
self-energy and retardation effects in a consistent way, is still lacking. 

We present here a strong coupling theory for the critical temperature $T_{\rm c}$ of $p$-wave pairing of spin polarised fermions in a BEC. Including 
the full frequency and momentum dependence of the induced interaction between the fermions caused by the exchange of phonons in the 
BEC, we show that retardation as well as self-energy effects can significantly suppress $T_{\rm c}$. We perform a systematic analysis varying both the
boson-boson and boson-fermion interaction strengths as well as their mass ratio, with an emphasis on experimentally relevant atomic mixtures.  
 This allows us to determine the most suitable systems and the optimal  conditions for which  $p$-wave superfluidity due to an induced 
 interaction   can be realised.

\paragraph{Model.--}
We consider a three-dimensional system consisting of spin-polarized, non-interacting fermions of mass $m_{\rm F}$ and density $n_{\rm F}$, mixed with bosons of mass $m_{\rm B}$ and density $n_{\rm B}$. The Bose gas is weakly interacting so that it can be described by  Bogoliubov theory well below the critical temperature for 
Bose-Einstein condensation. At temperature $T = \beta^{-1}$ ($\hbar = k_{\rm B} = 1$), the 
  properties of the mixture are described by the partition function  
\beq
\calZ = \int  \calD(\bar a, a) \int \calD (\gamma^*,\gamma) e^{-(S^0_{\rm F}+S_{\rm B} + S_{\rm int})},
\label{pf}
\eeq
where $(a,\bar a)$ and $(\gamma,\gamma^*)$  are Grassmann and complex fields
for the fermions and Bogoliubov phonons respectively. 
The  action for the free fermions is 
\beq
S_{\rm F}^0 =  \sum_{\bp ,n}  \bar a(p)\left (-i\omega_n + \xi_\bp \right )a(p),
\eeq
 where $p \equiv (\bp, i\omega_n)$, $\omega_n =(2n+1)\pi T$ is a Fermi Matsubara frequency, and $\xi_\bp = {\bp^2}/{2m_{\rm F}} -\mu_{\rm F}$ is the free fermion dispersion measured from the chemical potential $\mu_{\rm F}$ of the Fermi gas. The action for the Bose gas is given by
\begin{align}
S_{\rm B} =  \sum_{\bq\neq0,\nu}  \gamma^*(q)(-i\omega_\nu+E_\bq)\gamma(q),
\end{align}
where $q \equiv (\bq, i\omega_\nu)$, $\omega_\nu =2\nu\pi T$ is a Bose Matsubara frequency, and 
$E_{\bq}  =\sqrt{ \varepsilon_{\bq} (\varepsilon_{\bq} + 2 g_{\rm B}n_{\rm B}) }$
is the Bogoliubov spectrum. Here $\varepsilon_{\bq} = \bq^2/2m_{\rm B}$ and $g_{\rm B} = 4\pi a_{\rm B}/m_{\rm B}$, where $a_{\rm B}$ is the boson scattering length.  Finally the fermion-boson interaction is 
\begin{align}
S_{\rm int}=  {g} \sqrt{\frac{n_{\rm B}}{\beta\calV}}\sum_{ \bq\neq 0,\nu}\sqrt{\frac{\varepsilon_{\bq}}{E_\bq}}\left [\gamma^*(q) + \gamma(-q) \right ] \rho(q),
\label{Interaction}
\end{align}
where $\calV$ is the system volume,  
$
\rho(\bq,i\omega_\nu) \equiv \sum_{\bp',n} \bar a(\bp'-\bq,i\omega_n-i\omega_\nu) a (\bp',i\omega_n)
$,
and $g = 2\pi a_{\rm FB}/m_{\rm r}$ is the boson-fermion interaction. Here $m_{\rm r} = m_{\rm F}m_{\rm B}/(m_{\rm F}+m_{\rm B})$ is the reduced mass and $a_{\rm FB}$ is the fermion-boson scattering length. In \eqref{Interaction} we did not include terms describing the  scattering between fermions and uncondensed bosons. Such terms can be neglected for the relatively weak boson-fermion interactions considered here~\cite{Rath2013,Christensen2015}, i.e.,   $k_{\rm F}|a_{\rm FB}|\lesssim 1$. 

The Bogoliubov  fields in Eq.~(\ref{pf}) can be integrated out, yielding an effective action for the fermions~\cite{Bardeen1966,Heiselberg2000,Bijlsma2000}  
\begin{align}
S_{\rm F}(\bar a, a) = S_{\rm F}^0(\bar a, a) +\frac{1}{2\beta\calV}\sum_{\bq, \nu}V_{\rm ind}(q)\bar\rho(q) \rho(q),
\label{SF}
\end{align}
where  $V_{\rm ind}$ is the phonon-mediated interaction given by
\beq
V_{\rm ind}(\bq,i\omega_\nu) = g^2 \frac{n_{\rm B}}{m_{\rm B}}\frac{\bq^2}{(i\omega_\nu)^2-E_\bq^2}. 
\label{Vinddef}
\eeq
This interaction corresponds to the exchange of one Bogoliubov mode between the fermions, treating the boson-fermion scattering as energy
 independent, which is valid for $k_{\rm F}|a_{\rm FB}|\lesssim1$. In the static case $\omega_\nu=0$,  Eq.~\eqref{Vinddef} is the Fourier transform of the well-known Yukawa 
 interaction with a range  given by the BEC coherence length $\xi_{\rm B}=1/\sqrt{8\pi n_{\rm B}a_{\rm B}}$. 
 
\paragraph{Eliashberg theory.--}
We investigate  pairing between fermions due to the mediated interaction \eqref{Vinddef}, focusing on 
reaching a high critical temperature $T_{\rm c}$. To describe 
such strong pairing in a reliable way,  we use  the Eliashberg theory retaining the full energy/momentum dependence of the normal and anomalous 
self-energies.  This framework has proven accurate for strong coupling electronic superconductors where the pairing is mediated by  phonons~\cite{Mahan2000book}, which is quite similar to the case at hand. 

\begin{figure}
  \includegraphics[width=0.99\columnwidth]{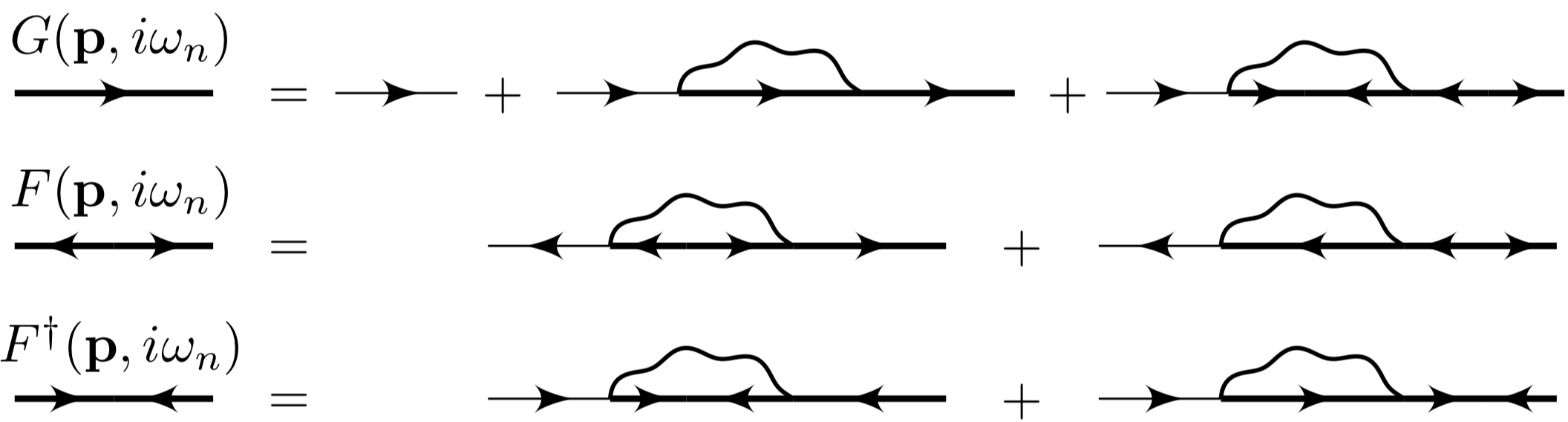}
\caption{Diagrammatic structure of the Eliashberg theory. The thin line represents the non-interacting  Green's function $G_0(\bp,i\omega_n) = 1/(i\omega_n - \xi_\bp)$ and the wavy curve represents the mediated interaction $V_{\rm ind}(\bq,i\omega_\nu)$.} 
  \label{fig:FeynmanFig}
\end{figure}

Eliashberg theory determines the normal and anomalous Green's functions, defined as $G(p) \equiv -\la a(p)\bar a(p)\ra$, $F(p) \equiv -\la a(p) a(-p)\ra$, and 
$F^\dagger(p) \equiv -\la \bar a(-p) \bar a(p)\ra$, where 
the expectation values are time-ordered. The Green's functions obey a generalised Dyson equation shown  diagrammatically in Fig.~\ref{fig:FeynmanFig}, which 
is solved by  
\begin{align}
G(p)  = \frac{i\omega_n + \xi_\bp + \Sigma (-p)}{[i\omega_n-A(p)]^2 -\left [\xi_\bp + S (p)\right ]^2 - |\Delta(p)|^2 }
\label{Gsol}
\end{align}
and 
\begin{align}
F(p)  = \frac{\Delta(p)}{[i\omega_n-A(p)]^2 -\left[\xi_\bp + S  (p)\right ]^2 - |\Delta(p)|^2},
\label{Fsol}
\end{align}
 with $F^\dagger(\bp,i\omega_n)=F(\bp,-i\omega_n)^*$. Here $\Sigma(p)$ is the normal self-energy, where $S(p) = [\Sigma(p)+\Sigma(-p)]/2$ and $A(p) = [\Sigma(p)-\Sigma(-p)]/2$ are its real and imaginary parts,
 and $\Delta (p)$ is the anomalous self-energy. The latter is essentially a momentum and frequency dependent pairing gap. The self-energies are evaluated using a generalised 
 Hartree-Fock approximation illustrated in Fig.~\ref{fig:FeynmanFig}, where the Hartree term is absorbed into a redefinition of the chemical potential $\mu_{\rm F}$.
 This gives 
 \begin{align}
\Sigma (p) = - \frac{1}{\beta \calV}\sum_{p'}V_{\rm ind}(p-p') G(p')
\label{Sigma}
\end{align}
for the normal Fock  self-energy and
\begin{align}
\Delta (p) = - \frac{1}{\beta \calV}\sum_{p'}V_{\rm ind}(p-p') F(p'),
\label{Delta}
\end{align}
for the  anomalous Fock self-energy.  We solve these 
equations  self-consistently for fixed fermion density
\begin{align}
n_{\rm F} & = \frac{1}{\beta \calV} \sum_{\bp,n} G(\bp,i\omega_n) e^{i\omega_n 0^+}.
\label{neq1}
\end{align}
A derivation of the Eliashberg equations using the path integral is given in the Supplemental Material~\cite{SM}. 

The pairing gap must be odd in momentum due to the Pauli principle for identical  fermions, and it can therefore 
be expanded in spherical harmonics $Y_{lm}(\hat \bp)$ with  $l=1,3,\ldots$. Since $T_{\rm c}$  is determined from the linearized forms of Eqs.~(\ref{Sigma})-(\ref{neq1}) which 
do not couple different $(l,m)$ channels, we use the $(l,m) = (1,1)$ ($p$-wave) ansatz 
 $\Delta(\bp,i\omega_n) = \Delta_{11}(|\bp|,i\omega_n) Y_{11}(\hat \bp)$, as this yields the highest $T_{\rm c}$. 
 The normal self-energy  $\Sigma(p)$ is spherically symmetric at $T_{\rm c}$ where there is no pairing to break this symmetry, and so we can write $\Sigma(\bp,i\omega_n) = \Sigma_{00}(|\bp|,i\omega_n) Y_{00}(\hat \bp)$.

In practice, we determine $T_{\rm c}$ by first evaluating the normal self-energy self-consistently assuming no pairing. Then we iterate Eqs.~\eqref{Sigma}-\eqref{neq1}
with a finite but very small initial value of the gap function. A decreasing (increasing) gap function under iteration 
indicates that the given temperature is above (below) $T_{\rm c}$. The details of the numerical procedure  are given in the Supplemental Material~\cite{SM}.

Note that we neglect the effects of the fermions on the bosons and assume a  temperature well below the critical temperature 
of the BEC, so that it can be treated using $T=0$ Bogoliubov theory. This is accurate if the  boson density  is much larger than that 
of the fermions, which is often the case experimentally. The effects of a Fermi gas on a BEC were considered in Ref.~\cite{Kinnunen2015}.

\paragraph{Qualitative analysis.--}
There are four physical parameters that  can be independently controlled in this system, namely the Fermi-Bose mass ratio $\alpha \equiv m_{\rm F}/m_{\rm B}$, 
density ratio $n_{\rm B}/n_{\rm F}$, scattering length $a_{\rm BF}$, and the boson scattering length $a_{\rm B}$. The critical  temperature $T_{\rm c}$ is determined by three dimensionless quantities formed out of these four parameters. The first two are the strength and the range of the mediated interaction, which can be estimated by considering its zero frequency component
\begin{align}
 V_{\rm ind}( \bq, 0)  = -\lambda \frac{\e_{\rm F}/k_{\rm F}^3}{(\bq/k_{\rm F})^2 +2/(k_{\rm F}\xi_{\rm B})^2}.
\end{align}
Here,  $\epsilon_{\rm F}=k_{\rm F}^2/2m_{\rm F}$ is the Fermi energy of the system with $k_{\rm F}=(6\pi^2n_{\rm F})^{1/3}$. The dimensionless quantity  
\begin{align}
\lambda=\frac{16}3(k_{\rm F}a_{\rm FB})^2\frac{n_{\rm B}}{n_{\rm F}}(1+\alpha)(1+1/\alpha)
\label{Strength}
\end{align}
measures the strength while  
\begin{align}
k_{\rm F}\xi_{\rm B} = \frac{\sqrt{3\pi}}{2}\sqrt{\frac{n_F}{n_{\rm B}k_{\rm F} a_{\rm B}}}
\label{corr_length}
\end{align}
 characterises the range of the mediated interaction. It is intuitively clear that increasing the strength and range of the pairing interaction will raise $T_{\rm c}$. 
   The third dimensionless quantity is the ratio of the speed of sound in BEC $c_{\rm B} = \sqrt{g_{\rm B} n_{\rm B}/m_{\rm B}}$ and the Fermi velocity $v_{\rm F} = k_{\rm F}/m_{\rm F}$,
 \begin{align}
 \frac{c_{\rm B}}{v_{\rm F}} = \sqrt{\frac {2} {3\pi}}\alpha \sqrt{(n_{\rm B}/n_{\rm F})(k_{\rm F}a_{\rm B})}.
 \label{speed_ratio}
 \end{align}
The larger this ratio is, the smaller the effects of retardation will be, and the higher the $T_{\rm c}$ will become.

Now a few comments are in order. First, Eqs.~\eqref{Strength}, \eqref{corr_length}, and \eqref{speed_ratio}
show that when the mass ratio $\alpha$ increases, the interaction strength
increases and its range is constant, while retardation effects decrease. 
This indicates that using a mixture of light bosons and heavy fermions favors a high $T_{\rm c}$, which we shall 
demonstrate explicitly below. 
Second, increasing $n_{\rm B}/n_{\rm F}$ will increase the speed  of sound in the BEC and the 
interaction strength,  but decrease its  range. 
Likewise, increasing   $k_{\rm F} a_{\rm B}$ will  increase the BEC 
speed  of sound but decrease the interaction 
range. The competition between these effects makes the dependence of $T_{\rm c}$ on $n_{\rm B}/n_{\rm F}$ and $k_{\rm F} a_{\rm B}$  \emph{a priori} non-trivial.  Finally, we cannot freely increase  the scattering
 length $a_{\rm FB}$,  as the system will phase separate (collapse) for sufficiently positive (negative) $a_{\rm FB}$. Within mean-field theory, the condition for avoiding such instabilities is~\cite{Viverit2000}
\begin{align}
(k_{\rm F}a_{\rm FB})^2\le\frac{2\pi}{(1+\alpha)(1+1/\alpha)}k_{\rm F}a_{\rm B}.
\label{PhaseSeparation}
\end{align} 
We emphasize however, that  \eqref{PhaseSeparation} most likely  underestimates the region of stability for trapped Bose-Fermi mixtures, since it is based 
on mean-field theory and  is  derived for a homogeneous system ignoring finite size effects. Indeed,  two recent experiments show that trapped 
Bose-Fermi mixtures are stable far beyond the condition given by \eqref{PhaseSeparation}, both for attractive~\cite{DeSalvo2017} and repulsive
 interactions~\cite{Lous2018}. Phase separation for trapped Bose-Fermi mixtures was considered in Refs.~\cite{Molmer1998,Nygaard1999,Roth2002}.

\paragraph{Numerical results.--}
We  now present numerical results for $T_{\rm c}$ for experimentally relevant  Bose-Fermi mixtures.
Since the BEC density is typically much higher than that of the fermions, we take $n_{\rm B}/n_{\rm F}=5$
for all the calculations.

\begin{figure}[th]
  \includegraphics[width=0.99\columnwidth]{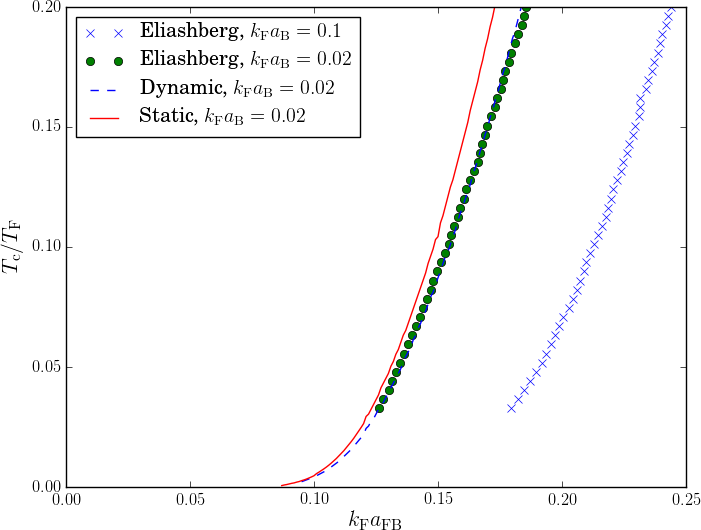}
  \caption{Critical temperature for $p$-wave pairing in the $^7$Li-$^{173}$Yb mixture as a function of Fermi-Bose scattering length $k_\mathrm{F}a_\mathrm{FB}$. The dynamic and static labels refer to when the normal self-energy, and to when both the 
  normal self-energy and retardation effects are ignored, respectively.}
  \label{fig:phase_diagram_ybli}
\end{figure}
Consider first the $^7$Li-$^{173}$Yb mixture, which has been experimentally realized~\cite{Schafer2018}. It corresponds to a mass ratio $\alpha=173/7$  as high as currently possible with present atomic gas experiments, and we expect it to be the most favorable for achieving a high $T_{\rm c}$. 
We plot in Fig.~\ref{fig:phase_diagram_ybli} the critical temperature  as a function of the  fermion-boson interaction strength $k_{\rm F} a_{\rm FB}$ obtained from the full 
Eliashberg theory using the boson-boson interaction strengths $k_{\rm F}a_{\rm B}=0.1$ and  $k_{\rm F}a_{\rm B}=0.02$. 
The critical temperature increases with $k_{\rm F} a_{\rm FB}$ as expected. 
Taking $T_{\rm c}/T_{\rm F}=0.1$ as a  conservative estimate for what can be realised experimentally, this is reached at the relatively weak coupling strengths 
$k_{\rm F}|a_{\rm FB}|\simeq 0.155$ for $k_{\rm F}a_{\rm B}=0.02$ and $k_{\rm F}|a_{\rm FB}|\simeq 0.215$ for $k_{\rm F}a_{\rm B}=0.1$, 
where our theory is reliable. The critical temperature is higher for $k_{\rm F}a_{\rm B}=0.02$ compared to  $k_{\rm F}a_{\rm B}=0.1$,  showing 
that when $k_{\rm F}a_{\rm B}$ decreases, the increase in  interaction range more than compensates for the increasing retardation effects. 

For comparison, we also plot in Fig.~\ref{fig:phase_diagram_ybli} the  critical temperature for $k_{\rm F}a_{\rm B}=0.02$ 
obtained when both retardation effects and the normal self-energy are neglected, i.e., we use the induced interaction evaluated at zero frequency and set $\Sigma (p)=0$.
Such a static theory significantly overestimates $T_{\rm c}$, which for this particular mixture is mainly because it neglects retardation effects. 
This can be seen when we include retardation but still neglect the normal self-energy $\Sigma (p)$, the resulting $T_{\rm c}$ largely agrees with that obtained 
from the full theory, as shown in Fig.~\ref{fig:phase_diagram_ybli}.  
Note that our results are independent of the sign of $a_{\rm FB}$, since the induced interaction is second order in $a_{\rm FB}$.

Equation \eqref{PhaseSeparation}  predicts that a homogeneous $^7$Li-$^{173}$Yb mixture  
will collapse/phase separate for $k_{\rm F}|a_{\rm FB}|\gtrsim 0.069$ for $k_{\rm F}a_{\rm B}=0.02$ and $k_{\rm F}|a_{\rm FB}|\gtrsim 0.153$ for $k_{\rm F}a_{\rm B}=0.1$. Taking into account that 
trapped mixtures are stable well beyond these critical values, we conclude from Fig.~\ref{fig:phase_diagram_ybli}
that a $^7$Li-$^{173}$Yb mixture is a promising platform for realizing a $p$-wave superfluid caused by an induced interaction. 

\begin{figure}[th]
  \includegraphics[width=0.99\columnwidth]{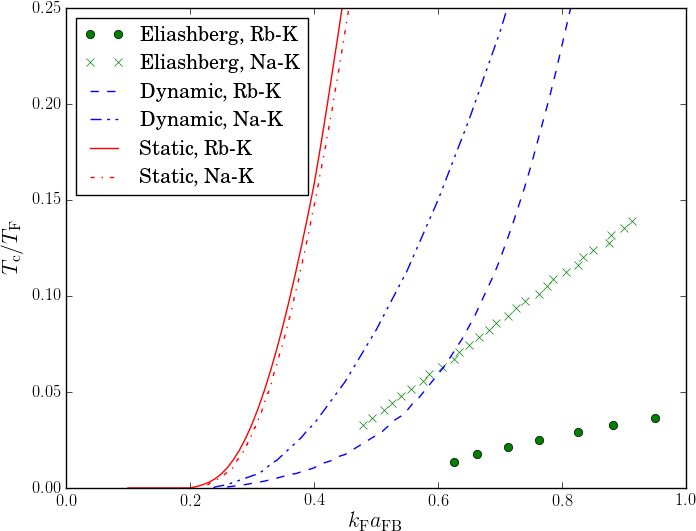}
  \caption{Critical temperature for the  $^{23}$Na-$^{40}$K and $^{87}$Rb-$^{40}$K mixtures as a function 
  of $k_{\rm F}a_{\rm FB}$ for $k_{\rm F}a_{\rm B}=0.02$. The static and dynamic labels 
  refer to the same theories as in Fig.~\ref{fig:phase_diagram_ybli}.}
  \label{fig:phase_diagram_nakrbk}
\end{figure}

Consider next the two experimentally relevant mixtures $^{23}$Na-$^{40}$K~\cite{Park2012} and $^{87}$Rb-$^{40}$K\cite{Zaccanti2006,Ospelkaus2006,Taglieber2008},
 which have almost the inverse mass ratios. 
In Fig.~\ref{fig:phase_diagram_nakrbk}, we plot $T_{\rm c}$ for $k_{\rm F}a_{\rm B}=0.02$
obtained using three theoretical approaches: i) the full Eliashberg theory, ii) including retardation but neglecting 
the normal self-energy, and iii) neglecting both retardation by using the zero frequency induced interaction and the normal self-energy.
The critical temperatures of the two mixtures are almost the same when both retardation and self-energy effects are ignored. This can be 
understood from Eq.~\eqref{Strength}, since the dimensionless interaction is nearly the same for the two mixtures. 
However,  $T_{\rm c}$ is much higher for the $^{23}$Na-$^{40}$K mixture, when retardation effects are included. 
This is because
 retardation is less important for light bosons due to their higher speed of sound, see Eq.~\eqref{speed_ratio}. 
Finally, Fig.~\ref{fig:phase_diagram_nakrbk}  shows that the normal self-energy $\Sigma (p)$  also suppresses $T_{\rm c}$ most
for the $^{87}$Rb-$^{40}$K mixture. The reason is that excitations in the BEC cost less energy for heavy bosons, 
which leads to larger self-energy effects. The fact that the 
$^{23}$Na-$^{40}$K mixture has a much higher $T_c$ than the 
 $^{87}$Rb-$^{40}$K mixture in the full Eliashberg theory nicely illustrates a main result of the present paper: a mixture of \emph{light bosons} and \emph{heavy fermions} is more favorable to achieve a high $T_{\rm c}$. This is further corroborated by the fact that 
according to Eq.~\eqref{PhaseSeparation}, the two mixtures become unstable almost at the same coupling 
strength, $k_{\rm F}|a_{\rm FB}|=0.171$ for $^{23}$Na-$^{40}$K and $k_{\rm F}|a_{\rm FB}|=0.165$ for $^{87}$Rb-$^{40}$K.

\begin{figure}[th]
  \includegraphics[width=0.99\columnwidth]{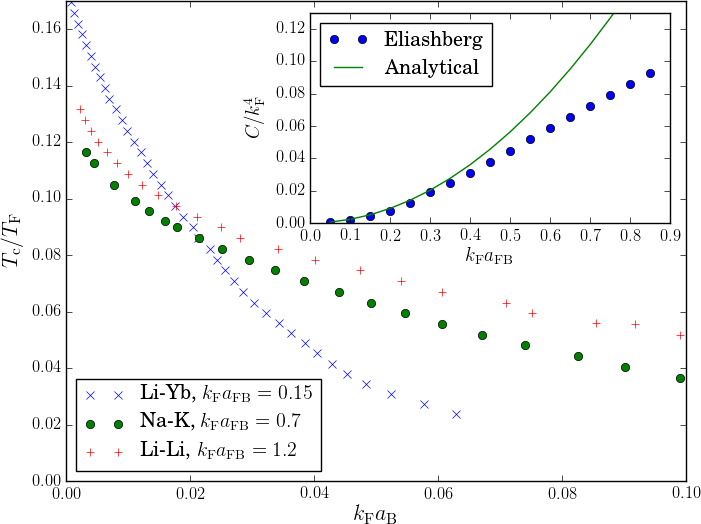}
  \caption{Critical temperature as function of the Bose-Bose scattering length $k_{\rm F}a_{\rm B}$ for three mixtures. Inset shows 
  the contact $C$ for the $^{23}$Na-$^{40}$K mixture.}
  \label{fig:phase_diagram_kfabsweep}
\end{figure}

In order to investigate the effects of the boson-boson interaction, we plot in Fig.~\ref{fig:phase_diagram_kfabsweep} the critical temperature as a function 
of $k_{\rm F}a_{\rm B}$ for the $^7$Li-$^{173}$Yb, $^{23}$Na-$^{40}$K, and $^7$Li-$^6$Li mixtures~\cite{Schafer2018,Park2012,Ferrier-Barbut2014}. 
For all three mixtures, $T_{\rm c}$ decreases with $k_{\rm F}a_{\rm B}$. Thus, although the sound velocity of the BEC increases with $k_{\rm F}a_{\rm B}$ 
thereby reducing retardation effects, this effect is overwhelmed by the corresponding reduction in the interaction range, so that the net effect is a suppression of $T_{\rm c}$ with increasing $k_{\rm F}a_{\rm B}$. The suppression is largest for the $^7$Li-$^{173}$Yb mixture, since retardation effects are already
small for light bosons so that a decrease in the interaction range has a larger relative effect. 

Finally, we plot in  the inset of Fig.~\ref{fig:phase_diagram_kfabsweep} the contact $C=\lim_{k\rightarrow\infty}\langle a_\bk^\dagger a_\bk\rangle\cdot k^4$~\cite{Tan2008a,Tan2008b,Braaten2012} for a $^{23}$Na-$^{40}$K mixture with $k_{\rm F}a_{\rm B}=0.1$ and temperature $T=0.125\,T_\mathrm{F}$. It increases with $k_{\rm F}a_{\rm BF}$ as expected. Since  
the Fock self-energy, Eq.~\eqref{Sigma} includes all dominant second order diagrams for a fermion
interacting with a BEC~\cite{Christensen2015,Camacho2018},  our theory  recovers the exact second order result~\cite{SM} 
\begin{align}
C=\frac{(2k_{\rm F}a_{\rm FB})^2n_{\rm B}}{9\pi^2 n_{\rm F}}k_{\rm F}^4.
\label{Contact}
\end{align}
We see from the inset of Fig.~\ref{fig:phase_diagram_kfabsweep} 
that the numerical results indeed approach Eq.\ \eqref{Contact} for $k_{\rm F}a_{\rm FB}\ll 1$ thereby illustrating the accuracy of our approach. 

\paragraph{Conclusions.--}
We presented a strong coupling theory for the $p$-wave pairing of spin polarised fermions in a BEC, which takes into account 
the full frequency and momentum dependence of the induced interaction between the fermions caused by the exchange of phonons in the 
BEC. Focusing on experimentally relevant systems, we calculated the critical temperature  varying the boson-boson and 
boson-fermion interaction strengths, as well as their mass ratio. Both retardation  as well as self-energy effects were shown to  significantly 
affect $T_{\rm c}$. Our systematic analysis allowed us to identity  the most suitable system 
for which the $p$-wave superfluidity can be achieved. In particular,  we showed that it is  within experimental 
reach using a mixture of light bosons and heavy fermions. The $p$-wave superfluid can be considered as the many-body limit 
of a gas of bi-polarons~\cite{Camacho2018b}, where the size of the bi-polarons is much larger than their average distance. This opens 
up the intriguing possibility to study the BEC-BCS crossover in an entirely new setting by varying the fermion density. 

\begin{acknowledgments}
 G.M.B.~wishes to acknowledge the support of the Villum Foundation and the  Danish Council of Independent Research.   The computational time was provided by the Triton cluster of the Aalto School
of Science.
\end{acknowledgments}

\bibliographystyle{apsrev4-1}
\bibliography{eliashberg}


\newpage

\onecolumngrid
\begin{center}
\newpage\textbf{
Supplemental Material\\[4mm]
\large Induced $p$-wave pairing in Bose-Fermi mixtures}
\\
\vspace{4mm}
{Jami J. Kinnunen$^{1}$, Zhigang Wu$^{2}$ and Georg M.~Bruun$^{3}$}\\
\vspace{2mm}
{\em \small
$^1$Department of Applied Physics, Aalto University, FI-00076 Aalto, Finland\\
$^2$Shenzhen Institute for Quantum Science and Engineering and Department of Physics, Southern University of Science and Technology, Shenzhen 518055, China\\
$^{3}$Department of Physics and Astronomy, Aarhus University, DK-8000 Aarhus C, Denmark\\}
\end{center}

\setcounter{equation}{0}
\setcounter{figure}{0}
\setcounter{table}{0}
\setcounter{section}{0}
\setcounter{page}{1}
\makeatletter
\renewcommand{\theequation}{S.\arabic{equation}}
\renewcommand{\thefigure}{S\arabic{figure}}
\renewcommand{\thetable}{S\arabic{table}}
\renewcommand{\thesection}{S.\arabic{section}}

\section{I. Derivation of Eliashberg equations in path integral formulism }
The Eliashberg equations are usually derived by means of the equation of motion. Here we present a diagrammatic derivation in the path integral formalism which, to our knowledge, has not been done before in the literature. We begin with the partition function 
after integrating out the boson fields under the Bogoliubov approximation
\beq
\calZ = \int \calD(\bar a, a) e^{-S_{\rm F}(\bar a, a)},
\label{pf3}
\eeq
where $S_{\rm F}(\bar a, a) $ is the effective action for the fermions 
given by \eqref{SF}. The physical quantities of interest are the normal and anomalous Green's function 
$
 G(p)  = - \la  a(p) \bar a (p)$, 
 $ F(p)  = - \la  a(p)  a (-p)\ra $, and $F^\dagger(p) = -\la \bar a(-p) \bar a(p)\ra$,
where $\la\cdots\ra  \equiv \calZ^{-1}  \int  \calD(\bar a, a)  \cdots e^{-S_{F}(\bar a, a)} $. 
To determine these Green's functions in the path integral formalism, we first express the action as 
\begin{align}
S_{\rm F}(\bar a, a)  &= \sum_{p}  \bar a(p)\left (-i\omega_n +\xi_\bp\right )a(p)  +\frac{1}{2\beta \calV}\sum_{p,p'}V_{\rm ind}(p-p')\bar a(-p) \bar a(p) a(p') a (-p') \nn \\
&+ \frac{1}{2\beta \calV}\sum_{k\neq 0}\sum_{p,p'}V_{\rm ind}(p-p')\bar a(-p) \bar a(p+k) a(p'+k) a (-p'),
\end{align}
where we have separated the term responsible for the Cooper pairing from the rest of the interaction terms.  
Next we introduce a pairing field $\Xi(p)$ via the Hubbard-Stratonovich transformation and eliminate this term (the second term in $S_{\rm F}(\bar a, a)$) in favour of terms for which the gauge invariance is broken explicitly.  The Hubbard-Stratonovich transformation is the following identity
\begin{align}
 &{\rm{det}} \left (-\frac{V_{\rm ind}}{2\beta \calV} \right ) \exp \left \{ - \frac{1}{2\beta \calV} \sum_{p p'} V_{\rm ind}(p-p') \bar a(-p)\bar a(p) a(p')a(-p') \right \} \nn \\
 =&\int \calD(\Xi^*, \Xi) \exp \left \{ \frac{\beta \calV}{2}\sum_{p,p'}\Xi^*(p)V_{\rm ind}^{-1}(p-p')\Xi(p') +\frac{1}{2}\sum_p \left [ \Xi^*(p)a(p)a(-p)+\Xi(p) \bar a(-p) \bar a(p)\right ]\right \},
\label{HST}
\end{align} 
where $V_{\rm ind}^{-1}(p-p')$ is the inverse matrix of $V_{\rm ind}(p-p')$ such that $\sum_q V_{\rm ind}^{-1}(p-q)V_{\rm ind}(q-p') = \delta_{p p'}$. Using the above transformation in the partition function we obtain
 \begin{align}
 \calZ ={\rm{det} \left (-{2\beta \calV}V_{\rm ind}^{-1} \right ) } \int  \calD(\Xi^*, \Xi)  e^ { \frac{\beta \calV}{2}\sum_{p,p'}\Xi^*(p)V_{\rm ind}^{-1}(p-p')\Xi(p')} \int \calD (\bar a, a ) e^{ -S' ( \bar a, a;\Xi^*,\Xi) },
 \end{align}
 where 
 \begin{align}
 S' ( \bar a, a;\Xi^*,\Xi) &= \sum_{p} \left (-i\omega_n +\xi_\bp  \right )\bar a(p) a(p) - \frac{1}{2}\sum_p \left [ \Xi^*(p)a(p)a(-p)+\Xi(p) \bar a(-p) \bar a(p)\right ]  \nn \\
& + \frac{1}{2\beta \calV}\sum_{k\neq 0}\sum_{p,p'}V_{\rm ind}(p-p')\bar a(-p) \bar a(p+k) a(p'+k) a (-p').
\label{Sprime}
 \end{align}
 We see that $S'$ is an action which describes the fermions in the presence of the gauge-symmetry-breaking fields $\Xi$. 
 Letting  
\begin{align}
\calZ' (\Xi^*, \Xi)& =  \int \calD (\bar a, a ) e^{ -S' ( \bar a, a;\Xi^*,\Xi) },
\end{align} 
we can write 
 \begin{align}
 \calZ &= {\rm{det} \left (-{2\beta \calV}V_{\rm ind}^{-1} \right ) } \int  \calD(\Xi^*, \Xi)  \exp \left \{ \frac{\beta \calV}{2}\sum_{p,p'}\Xi^*(p)V_{\rm ind}^{-1}(p-p')\Xi(p') + \ln \calZ' (\Xi^*, \Xi)\right \} \nn \\
 & \equiv  \int  \calD(\Xi^*, \Xi)  e^{-S_{\rm eff}(\Xi^*,\Xi)},
 \end{align}
 where 
 \begin{align}
 S_{\rm eff}(\Xi^*,\Xi) =  -\frac{\beta \calV}{2}\sum_{p,p'}\Xi^*(p)V_{\rm ind}^{-1}(p-p')\Xi(p') - \ln \calZ' (\Xi^*, \Xi) + {\rm const.} 
 \label{Seff}
 \end{align}
The purpose of introducing the such auxiliary fields can be seen from the following formal identity
 \begin{align}
 \overline \Xi(p)
 &=  -\sum_{p'}V_{\rm ind}(p-p')\la a (p')a(-p') \ra =\sum_{p'}V_{\rm ind}(p-p') F(p') ,
 \label{XiF}
\end{align}
 where $\overline \cdots  \equiv  \calZ^{-1} \int  \calD(\Xi^*, \Xi)  \cdots e^{-S_{\rm eff}(\Xi^*,\Xi)}$.  This identity connects the anomalous Green's function to the average of the auxiliary $\Xi(p)$ field with respect to the effective action in Eq.~(\ref{Seff}). 
 
With the auxiliary pairing field, the calculation of the Green's functions now takes two steps. First, we evaluate the Green's functions with respect to the action $S'$. Namely we first determine
 \begin{align}
  \label{Gprime}
 G'(\bp,i\omega_n;\Xi) &= -  \calZ'^{-1} \int \calD (\bar a, a ) a(p) \bar a (p) e^{ -S' ( \bar a, a;\Xi^*,\Xi) } \equiv  -\la a(p) \bar a (p) \ra' ;\\
  F'(\bp,i\omega_n;\Xi) &=-  \calZ'^{-1} \int \calD (\bar a, a ) a(p)  a (-p) e^{ -S' ( \bar a, a;\Xi^*,\Xi) } \equiv  -\la a(p)  a (-p) \ra',
 \label{Fprime}
 \end{align}
 where $\la\cdots\ra'  \equiv \calZ'^{-1}  \int  \calD(\bar a, a)  \cdots e^{-S'(\bar a, a;\Xi,\Xi^*)} $.  
 Clearly these Green's functions are functions of the pairing field $\Xi$. 
 Next the Green's functions can be evaluated as 
  \begin{align}
  \label{G2}
 G(\bp,i\omega_n) =  \calZ^{-1} \int  \calD(\Xi^*, \Xi)  G'(\bp,i\omega_n;\Xi) e^{-S_{\rm eff}(\Xi^*,\Xi)} \approx  G'(\bp,i\omega_n;\overline \Xi);  \\
  F(\bp,i\omega_n) =  \calZ^{-1} \int  \calD(\Xi^*, \Xi)  F'(\bp,i\omega_n;\Xi) e^{-S_{\rm eff}(\Xi^*,\Xi)} \approx   F'(\bp,i\omega_n;\overline \Xi),
 \label{F2}
 \end{align}
 where $\la\cdots\ra_{\rm eff}  \equiv  \calZ^{-1} \int  \calD(\Xi^*, \Xi)  \cdots e^{-S_{\rm eff}(\Xi^*,\Xi)}$. In the last step, we have adopted a mean-field approximation. In doing so we have obtained three coupled equations (\ref{XiF}), (\ref{G2}) and (\ref{F2}), which can be solved self-consistently. 

\begin{figure}[b]
  \includegraphics[width=0.6\columnwidth]{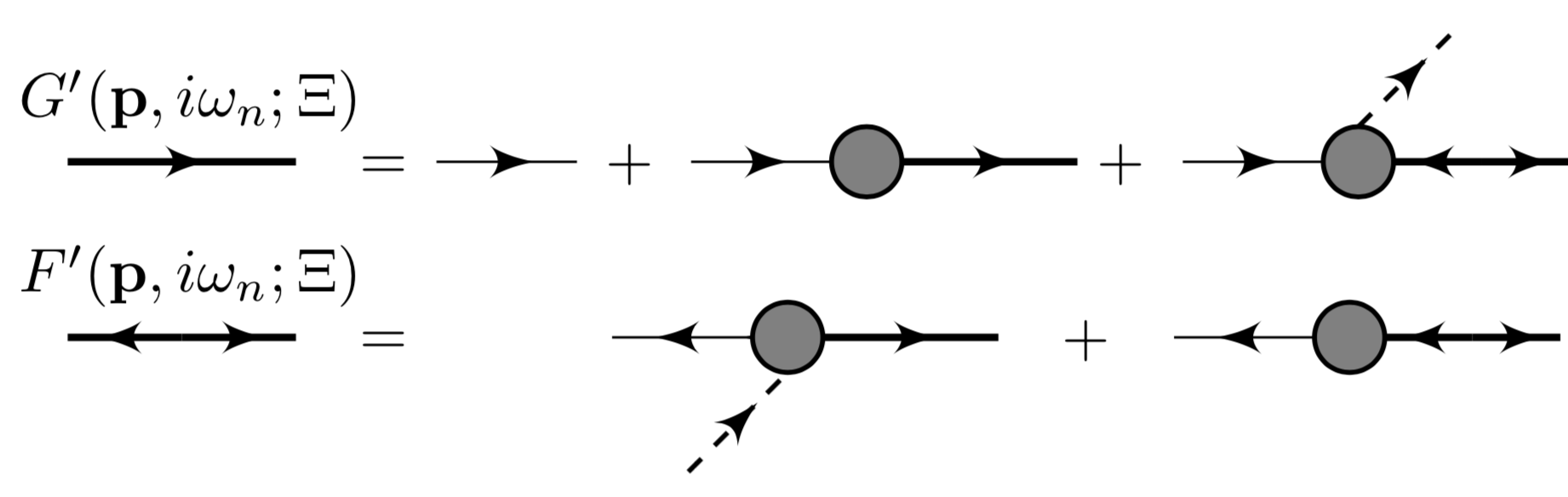}
  \caption{The diagrammatic representation of the \eqref{Gprimeeq} and \eqref{Fprimeeq}. The thin line represents the non-interacting Green's function $G_0(p)$, the dashed lines represent the pairing fields $\Xi$ and $\Xi^*$ and the circles represent the normal and anomalous self-energies.}
  \label{fig:GFD}
\end{figure}
\begin{figure}
 \includegraphics[width=0.6\columnwidth]{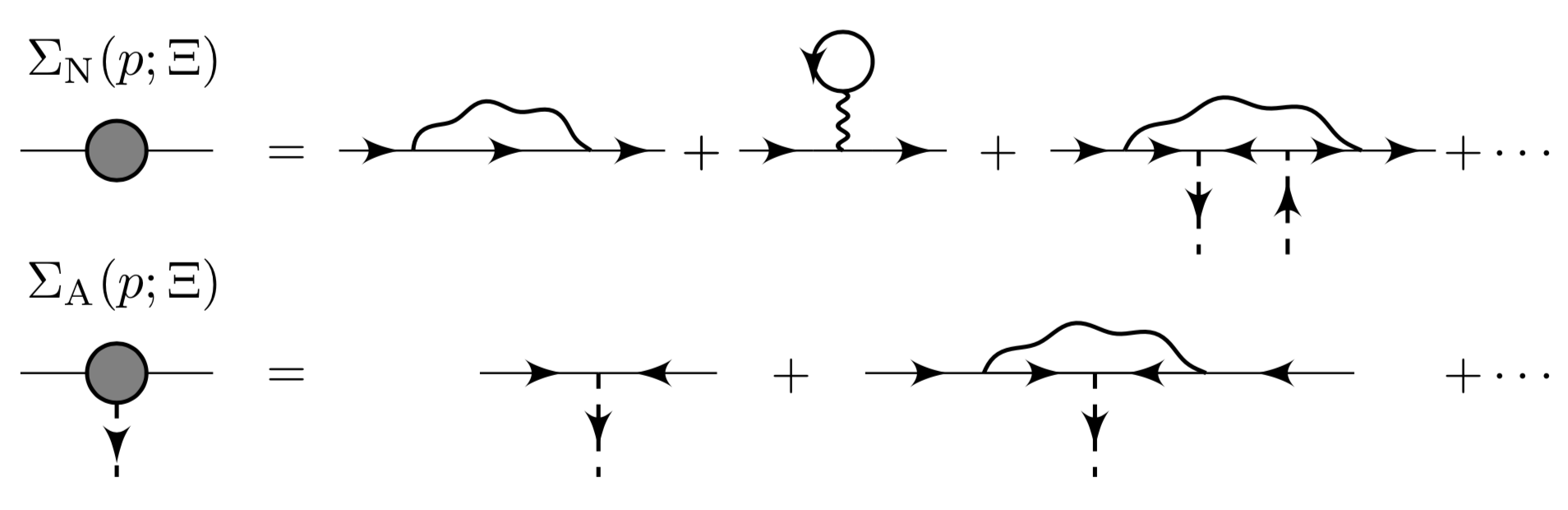}
  \caption{The first few diagrams for the normal and anomalous self-energies.}
  \label{fig:SFD}
\end{figure}
Now, we still need to determine the Green's functions $G'(\bp,i\omega_n;\Xi)$ and $F'(\bp,i\omega_n;\Xi)$ which are defined with respect to the action $S'$ in Eq.~(\ref{Sprime}).  Since the action $S'$ contains the pairing field which breaks the gauge symmetry explicitly, we can calculate the both $G'(\bp,i\omega_n;\Xi)$ and $ F'(\bp,i\omega_n;\Xi)$ by means of a standard diagrammatic method based on the Wick theorem, similar to the case of Bose gas in the presence of a condensate. For readers who are familiar with the diagrammatic perturbation theory of a Bose condensate, it is not difficult to show that the Green's functions defined in Eq.~(\ref{Gprime}) and (\ref{Fprime})  satisfy the following equations 
\begin{align}
\label{Gprimeeq}
G'(p;\Xi) = G_0(p) + G_0(p;\Xi) \Sigma'_{\rm N}(p;\Xi)G'(p;\Xi) + G_0(p;\Xi) \Sigma'_{\rm A}(p;\Xi) F^{\prime \dag }(p;\Xi);  \\
F^{\prime \dag}(p;\Xi) = G_0(-p) \Sigma'_{\rm N}(-p;\Xi)F^{\prime \dag}(p;\Xi) + G_0(-p) \Sigma^{\prime *}_{\rm A}(-p;\Xi) G'(p;\Xi),
\label{Fprimeeq}
\end{align}
where $\Sigma'_{\rm N}(p;\Xi)$ and $\Sigma'_{\rm A}(p;\Xi)$ are the normal and anomalous self energies respectively. 
The first few diagrams for the normal and anomalous self energies are shown in Fig.~\ref{fig:SFD}. 
The Eliashberg theory corresponds to evaluating $\Sigma_{\rm N}$ by the dressed Fock diagram and $\Sigma_{\rm A}$ by the diagram in the first order of $\Xi$. In this approximation, the latter is given by $\Sigma_{\rm A}(p,\Xi)  = \frac{1}{2}\left[ \Xi(p) - \Xi(-p) \right ]$. Since $\overline{\Xi}(-p) = - \overline{\Xi}(p) $, we find $\Sigma_{\rm A}(p,\overline\Xi)  = \overline \Xi(p)$. Now we can see that $\overline \Xi(p)$ is simply the gap parameter $\Delta(p)$ in the main text. Solving \eqref{Gprimeeq} and \eqref{Fprimeeq} with this in mind, we arrive at the \eqref{Gsol} and \eqref{Fsol} in the main text.

\section{II. Remarks on numerical solutions of the Eliashberg equations}
With the ansatz used, the equations we ultimately solve are
\begin{align}
\Delta_{11}(|\bp|,i\omega_n) &= -\frac{2\pi}{\beta} \sum_{n'}\int \frac{d\bp'}{(2\pi)^3}  V_1(i\omega_n-i\omega_{n'},|\bp|,|\bp'|) \nn \\
&\times \frac{Y^*_{11}(\theta',\phi')Y_{11}(\theta',\phi')\Delta_{11}(|\bp'|,i\omega_{n'})}{[\omega_{n'}-iA_{00}(|\bp'|,i\omega_{n'})Y_{00}(\theta',\phi')]^2 + \left [\xi_{\bp'} + S_{00} (|\bp'|,i\omega_{n'}) Y_{00}(\theta',\phi')\right ]^2 + |\Delta_{11}(|\bp'|,i\omega_{n'})Y_{11}(\theta',\phi')|^2 },
\label{gapeq2}
\end{align}
\begin{align}
\Sigma_{00}(|\bp|,i\omega_n) &= -\frac{2\pi}{\beta} \sum_{n'}\int \frac{d\bp'}{(2\pi)^3}  V_0(i\omega_n-i\omega_{n'},|\bp|,|\bp'|)  \nn \\
&\times \frac{Y^*_{00}(\theta',\phi')\left(i\omega_{n'} + \xi_{\bp'} e^{-i\omega_{n'} 0^+} + \Sigma_{00} (|\bp'|,-i\omega_{n'})Y_{00}(\theta',\phi')\right)}{[\omega_{n'}-iA_{00}(|\bp'|,i\omega_{n'})Y_{00}(\theta',\phi')]^2 +\left [\xi_{\bp'} + S_{00} (|\bp'|,i\omega_{n'}) Y_{00}(\theta',\phi')\right ]^2 + |\Delta_{11}(|\bp'|,i\omega_{n'})Y_{11}(\theta',\phi')|^2 },
\label{sf2}
\end{align}
and
\begin{align}
n_{\rm F} & \equiv \frac{1}{\beta \calV} \sum_{n'\bp'} G(\bp',i\omega_{n'}) e^{i\omega_{n'} 0^+} \nn \\ 
& = -\frac{1}{\beta} \sum_{n'}\int \frac{d\bp'}{(2\pi)^3} \frac{i\omega_{n'} e^{i\omega_{n'} 0^+}+ \xi_{\bp'}+ \Sigma_{00}(|\bp'|,-i\omega_{n'}) Y_{00}(\theta',\phi')}{[\omega_{n'}-iA_{00}(|\bp'|,i\omega_{n'})Y_{00}(\theta',\phi')]^2 +\left [\xi_{\bp'} + S_{00} (|\bp'|,i\omega_{n'}) Y_{00}(\theta',\phi')\right ]^2 + |\Delta_{11}(|\bp'|,i\omega_{n'})Y_{11}(\theta',\phi')|^2 },
\label{neq2}
\end{align}
where 
\begin{align}
V_l(i\omega_\nu,|\bp|,|\bp'|) &= \int_{-1}^1 d \cos(\phi-\phi') P_l (\cos(\phi-\phi')) V_{\rm ind}(i\omega_\nu,|\bp-\bp'|) \nn \\
& =-4g^2 m_{\rm B}n_{\rm B} \int_{-1}^1 d x P_l(x) \frac{|\bp|^2+|\bp'|^2-2|\bp||\bp'|x}{\left ( |\bp|^2+|\bp'|^2-2|\bp||\bp'|x + 2m_{\rm B}g_{\rm B}n_{\rm B}\right )^2 -4m_{\rm B}^2\left ( g_{\rm B}^2n_{\rm B}^2- \omega_\nu^2\right ) }.
\end{align}
Solving the equations~\eqref{gapeq2},~\eqref{sf2}, and~\eqref{neq2} self-consistently requires an iterative procedure. However, since we are only determining the critical temperature of the $p$-wave superfluid transition, the normal self-energy $\Sigma_{00}(|\bp|,i\omega_n)$ can be calculated self-consistently by setting the superfluid order parameter to be zero in \eqref{sf2}.  Real parts of the normal self-energy are typically very large, of the order of Fermi energy or higher, meaning that the chemical potential is strongly shifted from the noninteracting value. This requires the solution of the number equation~\eqref{neq2}. In practice, we solve the self-energy by starting with a non-interacting finite temperature system. At each iteration of the self-energy $\Sigma_{00}(|\bp|,i\omega_n)$, we increase the strength of the Bose-Fermi interaction slightly (typically in steps of $\delta (k_{\rm F} a_{\rm FB}) = 0.05$, although for large mass ratios a smaller step needs to be used), calculate the new self-energy profile using Eq.~\eqref{sf2}, solve the chemical potential by iterating the number equation~\eqref{neq2}, and then proceed to the next iteration until the target interaction strength $k_{\rm F}a_{\rm FB}$ is reached.

Iterative solution requires the tabulation of the whole two-dimensional self-energy profile $\Sigma_{00}(|\bp|,i\omega_n)$. Unfortunately, the Matsubara summations are slowly converging, requiring the tables to have very high cutoffs. We use frequency cutoffs of up to 100 000 points -- since we operate in Matsubara frequency space, the frequency grid is naturally discrete, with the actual cutoff frequency depending on temperature as in $\omega_n = \frac{2\pi (2n+1)}{\beta}$. Even the high frequency cutoff employed in the tabulation is not enough for obtaining good accuracy in the iteration. To avoid cutoff effects, we extrapolate the tabulated self-energy to arbitrarily high frequencies by employing the known asymptotic relation $\Sigma_{00}(|\bp|,i\omega_n) \sim \kappa \sqrt{\omega_n}$ for large $\omega_n$. The prefactor $\kappa$ is obtained by fitting the asymptotic relation to the high frequency part of the tabulated self-energy. The requirement to get a good fit for $\kappa$ ultimately determines the needed magnitude of the frequency cutoff.

The momentum cutoff is varied from $8\,k_\mathrm{F}$ up to $96\,k_\mathrm{F}$ using a grid of 1024 or 2048 points. Together with the frequency grid, the resulting two-dimensional tables have typically $10^8$ elements, which is prohibitively large for any practical calculation. To make the problem tractable, we use a custom-made two-dimensional adaptive tabulation, in which the tabulation does not have equidistant grid spacing, but rather spacing is decreased wherever the curvature of the tabulated function is large. In practice this allows us to do the full tabulation of the self-energy $\Sigma_{00}$ with the order of $10^4$ individually evaluated elements.

Finally, some comments regarding solving the chemical potential and the number equation. The number equation is converging only due to the convergence factor $e^{i\omega_{n'} 0^+}$. In order to avoid the need for any unphysical convergence factors, we instead solve for momentum perturbations from the ideal finite-temperature distribution
$n({\bf k}) = f(k)  + \delta n({\bf k})$, where $f(k)$ is the Fermi-Dirac distribution
\begin{equation}
   f(k) = \frac{1}{1+e^{\beta \xi_\bk}},
\end{equation}
where $\xi_\bk = \frac{\hbar^2 k^2}{2m_F} - \mu$.
The momentum perturbation is now
\begin{align}
\delta n({\bf k}) = \frac{1}{\beta} \sum_{n} \left[G(|{\bf k}|,i\omega_{n})-\frac{1}{i\omega_n - \xi_\bk}\right].
\label{eq:momdist}
\end{align}
Since the diagonal Green's functions $G(|{\bf k}|,i\omega_n)$ have the same asymptotic frequency dependence as the non-interacting Green's function $1/(i\omega_n - \xi_\bk)$, the Matsubara summation is convergent without any convergence factors. Solving the momentum distribution yields straight forwardly the number equation
\begin{equation}
  n_{\rm F} = \frac{1}{(2\pi)^3} \int d{\bf k} \, n({\bf k}).
\end{equation}
The chemical potential $\mu$ that yields the correct fermionic density $n_F$ is obtained by iterating the momentum distribution and the number equation.
 
\section{The Contact}

\begin{figure}
  \includegraphics[width=0.48\columnwidth]{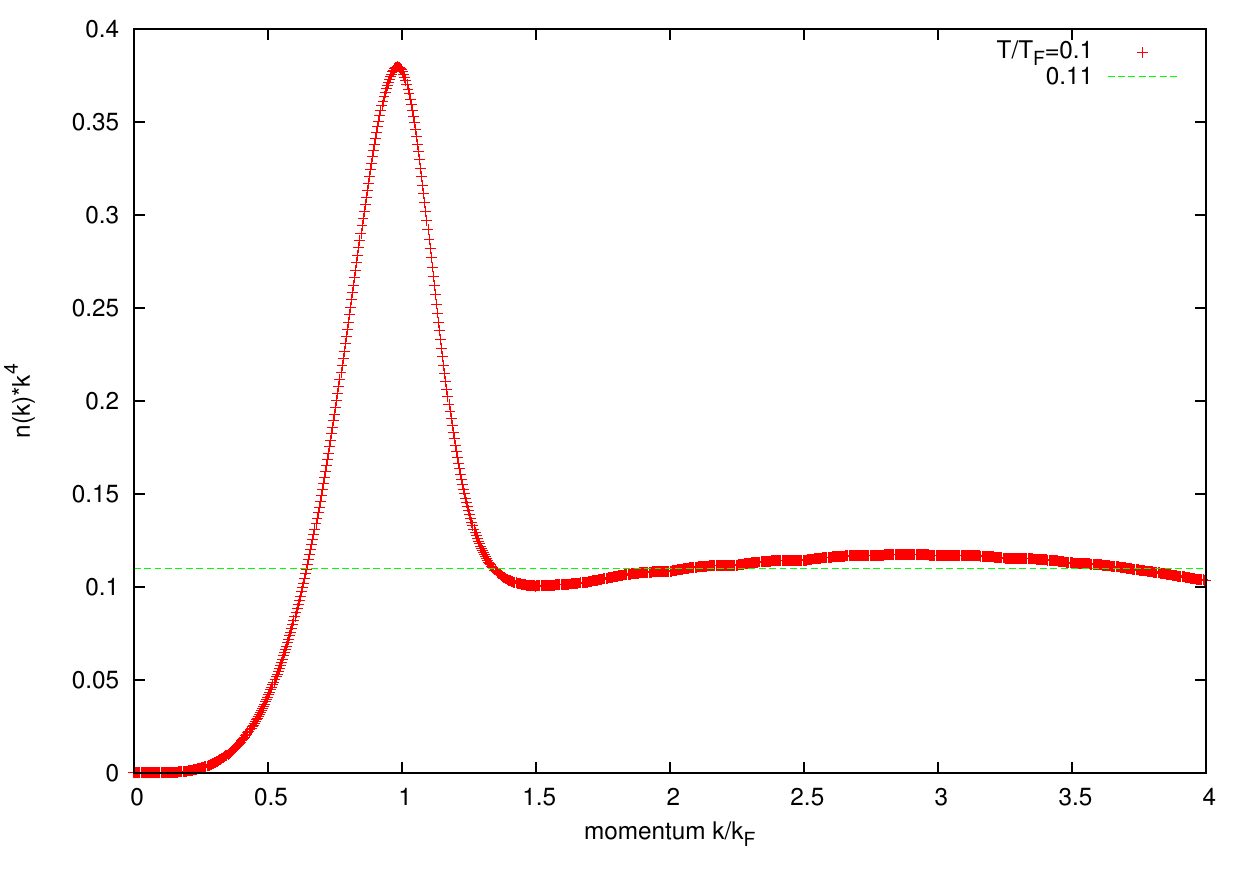}
  \caption{The momentum distribution of the fermionic gas for $^{23}$Na-$^{40}$K mixture at the superfluid phase transition ($T = 0.1 T_{\rm F}$, $n_{\rm B}=5n_{\rm F}$, $k_{\rm F}a_{\rm B} = 0.1$, $k_{\rm F}a_{\rm FB} = 1.00625$) scaled by factor $k^4$. Plot shows clearly the $1/k^4$ tail of the momentum distribution, with the roughly constant tail. The green line is a constant fit to the tail between momenta $1.5\,k_{\rm F}$ and $3.5\,k_{\rm F}$, yielding the contact parameter $C=0.11\,k_{\rm F}^4$.}
  \label{fig:contact}
\end{figure}
The contact $C$ can be defined in terms of the momentum distribution as 
\begin{equation} 
   C = \lim_{k\rightarrow \infty} n(k)k^4.
\end{equation}
Since we can calculate the momentum distributions numerically, we can also deduce the contact parameter due to the Fermi-Bose contact interaction.
Furthermore, since the asymptotic form of the momentum distribution $n(k) \sim k^{-4}$ is a general property of systems with contact interactions, we use it as one of the criteria for determining the numerical accuracy of the calculation.
In practice, the scaled momentum distribution $n(k)k^4$ should be only weakly dependent on the momentum for large $k$, in order to be able to obtain a good estimate of the contact parameter. As a typical example, see Fig.~\ref{fig:contact}.
The contact parameter can also be analytically calculated in the weakly interacting limit, which is a further check for the numerics. We now explain how the second order result \eqref{Contact}for the contact is obtained from our theory. First, we rewrite \eqref{neq1} in the usual way as 
\begin{align}
\langle a_{\bp}^\dagger a_\bp\rangle = \int_{-\infty}^\infty\!\frac{d\omega}{2\pi}\frac{e^{\omega\cdot0_+}}{e^{\beta\omega}+1}A(\bp,\omega), 
\label{neq3}
\end{align}
where $A(\bp,\omega)=-2\text{Im}G(\bp,\omega+i0_+)$ is the spectral function of the fermions. As explained in Ref.~\cite{Camacho2018}, the Fock diagram in Fig.~\ref{fig:FeynmanFig} with the induced interaction given by \eqref{Vinddef}, is identical to the second order self-energy of a fermion in a BEC.   
The second order retarded self-energy is  given by
\begin{align}
\Sigma(\bp,\omega)=-g^2n_\text{B}\int\!\frac{d^3q}{(2\pi)^3}\left[\frac{\epsilon_\bq}{E_\bq}\left(\frac{1+f_{\rm{B}}(E_\bq)+f(\xi_{\bp+\bq})}{E_\bq+\xi_{\bq+\bp}-\omega-i0_+}
+
\frac{f(\xi_{\bp+\bq})-f_{\rm{B}}(E_\bq)}{\omega+i0_++E_\bq-\xi_{\bq+\bp}}
\right)+\frac{2m_{\rm r}}{q^2}\right]
\label{Selfenergy}
\end{align}
where $f(x)=[\exp(\beta x)+1]^{-1}$ and $f_{\rm B}(x)=[\exp(\beta x)-1]^{-1}$ is the 
Fermi and Bose distribution function respectively~\cite{Christensen2015}. For large $|\bp|$, the Green's function 
has no real pole and the spectral function is then proportional to the imaginary part of the self-energy. The first term in Eq.~\eqref{Selfenergy} gives a contribution that is exponentially suppressed for large 
momenta, whereas the second term gives for $T=0$ and to second order in $g$
\begin{align}
\langle a_{\bp}^\dagger a_\bp\rangle=
g^2n_\text{B}\int\!\frac{d^3q}{(2\pi)^3}\frac{\epsilon_\bq}{E_\bq}
\frac{f(\xi_{\bp+\bq}-E_\bq)f(\xi_{\bq+\bp})}{(E_\bq-\xi_{\bq+\bp}+\xi_\bp)^2}.
\end{align}
This integral can easily be performed for large $|\bp|$ yielding Eq.~\eqref{Contact} in the main text.

\end{document}